\documentclass[nofootinbib,superscriptaddress,preprintnumbers]{revtex4}
\usepackage{amsmath}
\usepackage{amsfonts}
\usepackage{amssymb}
\usepackage{color}
\usepackage{graphicx}
\usepackage{graphics}
\usepackage{adjustbox,lipsum}
\usepackage[paperwidth=210mm,paperheight=297mm,centering,hmargin=2cm,vmargin=2.5cm]{geometry}
\def\beq{\begin{eqnarray}}
\def\eeq{\end{eqnarray}}
\def\bea{\begin{eqnarray}}
\def\eea{\end{eqnarray}}
\newcommand{\s}{\text{\tiny S}}

\newcommand{\dm}{\text{\tiny DM}}

\definecolor{blu}{cmyk}{1,0.7,0,0.6}
\begin{document}
\preprint{LPT-Orsay-15-29}

\title{\color{blu}A Model of Neutrino Mass \\
 and Dark Matter with an Accidental Symmetry}
\author{Amine Ahriche}
\email{aahriche@ictp.it}
\affiliation{Department of Physics, University of Jijel, PB 98 Ouled Aissa, DZ-18000
Jijel, Algeria.}
\affiliation{The Abdus Salam International Centre for Theoretical Physics, Strada
Costiera 11, I-34014, Trieste, Italy.}
\author{Kristian~L.~McDonald}
\email{klmcd@physics.usyd.edu.au}
\affiliation{School of Physics, The University of Sydney, NSW 2006, Australia.}
\author{Salah Nasri}
\email{snasri@uaeu.ac.ae}
\affiliation{Physics Department, UAE University, POB 17551, Al Ain, United Arab Emirates.}
\author{Takashi Toma}
\email{takashi.toma@th.u-psud.fr}
\affiliation{Laboratoire de Physique Th\'eorique, CNRS - UMR 8627,
 Universit\'e de Paris-Sud 11 F-91405 Orsay Cedex, France.}

\begin{abstract}
We present a model of radiative neutrino mass that automatically contains an
accidental $Z_2$ symmetry and thus provides a stable dark matter
candidate. This allows a common framework for the origin of neutrino
mass and dark matter without invoking any symmetries beyond those of
the Standard Model. The model can be probed by direct-detection
experiments and $\mu\rightarrow e+\gamma$ searches, and predicts a
charged scalar that can appear at the TeV scale, within reach of
collider experiments.
\vspace{0.3cm} 
\end{abstract}

\maketitle

\section{Introduction\label{sec:introduction}}

The existence of massive neutrinos provides concrete evidence for
physics beyond the Standard Model (SM). Similarly, the explanation
of observed galactic rotation curves in terms of gravitating dark
matter (DM) further suggests the SM is incomplete. Efforts to
explain these two key evidences for new physics are varied, though
an interesting approach is to seek a common or unified framework
that simultaneously solves both puzzles. For example, if small
neutrino masses are realized via radiative
effects~\cite{Zee:1980ai}, it is conceivable that DM plays a role in
generating the masses, allowing a type of unified description for
massive neutrinos and DM. This is the motivation for the models of
Krauss, Nasri and Trodden
(KNT)~\cite{Krauss:2002px,Baltz:2002we,JCAP-AN} and
Ma~\cite{Ma:2006km,Ho:2013hia}.
Both models extend the SM so that neutrino masses are generated
radiatively with DM propagating in the loop diagram. In
order to ensure DM stability (and preclude tree-level neutrino mass) a $%
Z_2$ symmetry is also imposed.

There are a number of generalizations of this basic idea which
similarly extend the SM to allow radiative neutrino mass via
couplings to
DM~\cite{Aoki:2008av,Kajiyama:2013zla,Ahriche:2014cda,Ahriche:2014oda,Ng:2014pqa,Culjak:2015qja}.
In common with the KNT and Ma models, the generalized models also
require the imposition of a new symmetry to render the DM
stable.\footnote{In some models the DM is merely sufficiently
long-lived, rather than absolutely stable. This does not require a
new symmetry but instead relies
on technically-natural parameter hierarchies (either among mass parameters~%
~\cite{Dev:2012bd} or dimensionless
couplings~\cite{Ahriche:2014oda}).} However, it is interesting to
consider models where DM stability instead results from an
accidental symmetry, in accordance with our experience from the SM,
where proton stability manifests the accidental baryon number
symmetry.

In this work we present a model of radiative neutrino mass that \emph{%
automatically} contains an accidental $Z_2$ symmetry and thus admits
a stable DM candidate. The model realizes a simple unified framework
for the origin of neutrino mass and DM while imposing only a minimal
symmetry structure, namely that of the SM. Neutrino mass appears at
the three-loop level via a diagram with the same topology as the KNT
model, while the DM is a neutral fermion with a non-trivial charge
under the accidental $Z_2$ symmetry. The model requires heavy DM
($M_\dm \sim 20$~TeV) and may be probed via DM direct-detection
experiments and future $\mu\rightarrow e+\gamma$ searches. It also
predicts a charged scalar that can appear at the TeV scale.

The layout of this paper is as follows. The model is introduced in
Section~\ref{sec:model7}. We calculate neutrino masses and discuss
important constraints in Section~\ref{sec:nuetrino_mass7}. Relevant
information regarding the DM is discussed in
Section~\ref{sec:dark_matter5} while our main numerical analysis and
results appear in Section~\ref{sec:results7}. Conclusions are drawn
in Section~\ref{sec:conc7}.

\section{The Model\label{sec:model7}}

\subsection{Field Content}

We extend the SM to include a charged singlet scalar, $S^{+}\sim (1,1,2)$, a
scalar septuplet, $\phi \sim (1,7,2)$, and three real
septuplet fermions, $\mathcal{F}_{i}\sim (1,7,0)$, where $i=1,2,3,$ labels
generations. We adopt the symmetric-matrix notation for the septuplets,
writing the scalar as $\phi _{abcdef}$, with $a,b,\ldots \in \{1,2\}$. The
components are given by
\begin{eqnarray}
&&\ \phi _{111111}=\phi ^{++++},\ \phi _{111112}=\frac{\phi ^{+++}}{\sqrt{6}}%
,\ \phi _{111122}=\frac{\phi ^{++}}{\sqrt{15}},\ \phi _{111222}=\frac{\phi
^{+}}{\sqrt{20}},\ \phi _{112222}=\frac{\phi ^{0}}{\sqrt{15}}, \notag \\
&&\ \phi _{122222}=\frac{\phi ^{-}}{\sqrt{6}},\ \phi _{222222}=\phi ^{--}\,,
\label{scalar_components7}
\end{eqnarray}%
where $\phi ^{++}$ and $\phi ^{--}$ are distinct fields, $\phi
^{--}\neq (\phi ^{++})^{*}$, and similarly $\phi ^{-}\neq (\phi ^{+})^{*}$. For the septuplet fermions, denoted as $\mathcal{F}_{abcdef}$, we have
\begin{eqnarray}
&&\ \mathcal{F}_{111111}=\mathcal{F}_{L}^{+++},\ \mathcal{F}_{111112}=\frac{%
\mathcal{F}_{L}^{++}}{\sqrt{6}},\ \mathcal{F}_{111122}=\frac{\mathcal{F}%
_{L}^{+}}{\sqrt{15}},\ \mathcal{F}_{111222}=\frac{\mathcal{F}_{L}^{0}}{\sqrt{%
20}},\ \mathcal{F}_{112222}=\frac{(\mathcal{F}_{R}^{+})^{c}}{\sqrt{15}},
\notag \\
&&\ \mathcal{F}_{122222}=\frac{(\mathcal{F}_{R}^{++})^{c}}{\sqrt{6}},\
\mathcal{F}_{222222}=(\mathcal{F}_{R}^{+++})^{c}.
\label{fermion_components7}
\end{eqnarray}%
The superscript \textquotedblleft $c$" denotes charge conjugation and the
numerical factors ensure the kinetic terms are canonically normalized. With
these fields, the Lagrangian contains the terms
\begin{equation}
\mathcal{L}\supset \mathcal{L}_{\text{{\tiny SM}}}-\;\frac{1}{2}\,\overline{%
\mathcal{F}_{i}^{c}}\,\mathcal{M}_{ij}\,\mathcal{F}_{j}\;+\{g_{i\alpha }\,%
\overline{\mathcal{F}_{i}}\,\phi \,e_{\alpha R}+f_{\alpha \beta }\,\overline{%
L_{\alpha }^{c}}\,L_{\beta }\,S^{+}+\mathrm{H.c.}\}\;-\;V(H,S,\phi ),
\label{L}
\end{equation}%
where lepton flavors are labeled by lower-case Greek letters,
$\alpha, \,\beta \in \{e,\,\mu,\,\tau \}$, and $L$ ($e_{R}$) is a SM
lepton doublet (singlet). The scalar potential is denoted as
$V(H,S,\phi )$. Note that the exotics $\phi $ and $\mathcal{F}$ do
not couple directly to the SM neutrinos, though they shall play a
key role in generating neutrino mass.

The explicit expansion for the fermion mass term is:
\begin{eqnarray}
&&-\frac{1}{2}\,(\overline{\mathcal{F}_{i}^{c}})_{abcdef}\,\mathcal{M}%
_{ij}\,(\mathcal{F}_{j})_{pqrstu}\,\epsilon ^{ap}\,\epsilon
^{bq}\,\epsilon ^{cr}\,\epsilon ^{ds}\,\epsilon ^{et}\,\epsilon
^{fu}+\mathrm{H.c.} \notag
\\
&=&-\mathcal{M}_{ij}\left\{ \overline{\mathcal{F}_{iR}^{+++}}\,\mathcal{F}%
_{jL}^{+++}-\overline{\mathcal{F}_{iR}^{++}}\,\mathcal{F}_{jL}^{++}+%
\overline{\mathcal{F}_{iR}^{+}}\,\mathcal{F}_{jL}^{+}-\frac{1}{2}\overline{(%
\mathcal{F}_{iL}^{0})^{c}}\,\mathcal{F}_{jL}^{0}\right\} +\mathrm{H.c.}
\notag \\
&=&-\mathcal{M}_{ij}\left\{ \overline{\mathcal{F}_{i}^{+++}}\,\mathcal{F}%
_{j}^{+++}+\overline{\mathcal{F}_{i}^{++}}\,\mathcal{F}_{j}^{++}+\overline{%
\mathcal{F}_{i}^{+}}\,\mathcal{F}_{j}^{+}+\frac{1}{2}\overline{\mathcal{F}%
_{i}^{0}}\,\mathcal{F}_{j}^{0}\right\},
\end{eqnarray}%
where we defined:
\begin{equation}
\mathcal{F}^{+++}=\mathcal{F}_{L}^{+++}+\mathcal{F}_{R}^{+++}\,,\quad
\mathcal{F}^{++}=\mathcal{F}_{L}^{++}-\mathcal{F}_{R}^{++}\,,\quad \mathcal{F%
}^{+}=\mathcal{F}_{L}^{+}+\mathcal{F}_{R}^{+}\,,\quad \mathcal{F}^{0}=%
\mathcal{F}_{L}^{0}-(\mathcal{F}_{L}^{0})^{c}.
\label{fermi7-mass_States}
\end{equation}%
Clearly $\mathcal{F}^{0}$ is a Majorana fermion, while the other six
components of $\mathcal{F}$ partner-up to give three massive charged
fermions (per generation). Without loss of generality, we choose a diagonal
basis for the fermions, such that $\mathcal{M}_{ij}=\mathrm{diag}%
(M_{1},\,M_{2},\,M_{3})$, with the masses ordered as $M_{1}<M_{2}<M_{3}$. We
shall see below that $\mathcal{F}$ does not mix with the SM leptons, to all
orders of perturbation theory, so Eq.~\eqref{fermi7-mass_States} describes
the mass eigenstates, which should be used in the Yukawa terms in Eq.~\eqref{L}. The lightest neutral fermion will play the role of DM,
and we denote its mass as $M_{\text{{\tiny DM}}}\equiv M_{1}$.

\subsection{An Accidental Symmetry}

The model contains an exact accidental $Z_{2}$ symmetry with action:
\begin{equation}
\{\phi,\,\mathcal{F}\}\rightarrow \{-\phi,\,-\mathcal{F}\}.
\label{eq:Z2_symmetry}
\end{equation}%
To see this, note that the potential can be written as
\begin{equation}
V(H,\,S,\,\phi )=V(H)+V(\phi )+V(S)+V_{m}(H,\,S)+V_{m}(H,\,\phi
)+V_{m}(S,\,\phi ).
\end{equation}%
The first four terms trivially preserve the discrete symmetry, while the
explicit forms for the last two mixing potentials are\footnote{%
The second term is equivalent to the standard $(H^{\dagger }\tau _{i}H){(%
}\phi ^{\dagger }T_{i}\phi )$ term, where $\tau _{i}$ and $T_{i}$ denote
$SU(2)$ generators for the distinct representations.}
\begin{equation}
V_{m}(H,\,\phi )=\lambda _{H\phi 1}(H^{\ast })^{a^{\prime }}H_{a^{\prime
}}(\phi ^{\ast })^{abcdef}\phi _{abcdef}+\lambda _{H\phi 2}(H^{\ast
})^{a^{\prime }}H_{a}(\phi ^{\ast })^{abcdef}\phi _{a^{\prime }bcdef},
\end{equation}%
and
\begin{equation}
V_{m}(S,\,\phi )=\lambda _{\text{{\tiny S}}\phi }|S|^{2}(\phi ^{\ast
})^{abcdef}\phi _{abcdef}+\frac{\lambda _{\text{{\tiny S}}}}{4}%
(S^{-})^{2}\phi _{abcdef}\phi _{a^{\prime }b^{\prime }c^{\prime }d^{\prime
}e^{\prime }f^{\prime }}\epsilon ^{aa^{\prime }}\epsilon ^{bb^{\prime
}}\epsilon ^{cc^{\prime }}\epsilon ^{dd^{\prime }}\epsilon ^{ee^{\prime
}}\epsilon ^{ff^{\prime }}+\mathrm{H.c.}
\end{equation}%
These potentials also preserve the symmetry defined by
Eq.~\eqref{eq:Z2_symmetry}. Note that there appears to be a third
distinct way to contract the $SU(2)$ indices in the mixing potential
$V_{m}(S,\,\phi ) $, namely
\begin{equation}
S^{-}(\phi ^{\ast })^{abcdef}\phi _{abca^{\prime }b^{\prime }c^{\prime
}}\phi _{defd^{\prime }e^{\prime }f^{\prime }}\epsilon ^{a^{\prime
}d^{\prime }}\epsilon ^{b^{\prime }e^{\prime }}\epsilon ^{c^{\prime
}f^{\prime }}.
\end{equation}
This would explicitly break the $Z_{2}$ symmetry. However, this term
is odd under the simultaneous interchange of the sets of dummy
indices $\{a,b,c\}\leftrightarrow \{d,e,f\}$ and
$\{a^{\prime},b^{\prime },c^{\prime }\}\leftrightarrow \{d^{\prime
},e^{\prime },f^{\prime }\}$~\cite{Ahriche:2014oda}, and thus
vanishes identically. The full theory therefore preserves the
accidental $Z_{2}$ symmetry defined by Eq.~\eqref{eq:Z2_symmetry}
and the model \emph{automatically} contains an absolutely stable
particle that is a DM candidate. The $Z_{2}$ symmetry also prevents
mixing between $\mathcal{F}$ and the SM leptons. To the best of our
knowledge this is the first such model of radiative neutrino mass
with an accidental symmetry that automatically gives a DM candidate.

At tree-level the components of $\mathcal{F}$ are mass-degenerate,
while the components of $\phi $ experience a mild splitting due to
the $\lambda _{H\phi 2}$-term in $V_{m}(H,\,\phi )$. For $M_{\phi
}\gtrsim \mathcal{O}(\mathrm{TeV})$ this mass-splitting is not
significant and is essentially negligible for $\lambda _{H\phi
2}\lesssim 0.1$. Thus, to good approximation the components of
$\mathcal{F}_i$ are degenerate at tree-level, with masses $M_{i}$,
as are the components of $\phi $ (with masses $M_{\phi }$).
Radiative corrections remove these mass degeneracies; loops
containing SM gauge bosons give small mass-splittings for the
components of $\mathcal{F}$, leaving $\mathcal{F}^{0}$ as the
lightest exotic fermion. Similar splittings are induced for the
components of $\phi $ which are readily
calculated with the results of Ref.~\cite{Cirelli:2005uq}. For most
purposes in this work these tiny splittings can be ignored.

The model contains two distinct $Z_2$-odd DM candidates, namely
$\mathcal{F}^0_1$ and $\phi^0$. However, $\phi^0$ has degenerate
real and imaginary components and also couples to the $Z$ boson.
This leads to tree-level $Z$ boson exchanges that are incompatible
with direct detection constraints. Thus, $\phi^0$ can be excluded as
a DM candidate, leaving $\mathcal{F}_1^0$ as the sole DM candidate
in the model and restricting one to the parameter space with
$M_\text{{\tiny DM}}=M_1<M_\phi$. The SM Higgs develops a nonzero
vacuum value, $\langle H\rangle\ne0$, breaking the electroweak
symmetry in the usual way. Furthermore, in the parameter space with
$\langle \phi \rangle=0$, which preserves the discrete symmetry, the
$\rho$-parameter retains its standard tree-level value.\footnote{We
shall see below that the septuplets must be heavier than the TeV
scale; given the very small mass-splittings, relative to the weak
scale, this should ensure that the new contributions
 to the oblique parameters are negligible. Also, similar to other models
with large multiplets, the $SU(2)_L$ coupling constant encounters a
Landau pole in the UV, due to the heavy septuplets.}

 Note that the inclusion of non-renormalizable
dimension 5 operators ($D=5$) can break the accidental symmetry. In particular, the $D=5$  operator
$HH\phi^\dag\phi^\dag\phi$  would allow the DM to decay. This feature is not specific to the present model; one expects global symmetries to be broken by gravitational effects~\cite{Witten:2000dt} so  non-renormalizable operators will, in general,  break global symmetries. This is true even in related models which impose e.g.~a global $Z_2$ or $U(1)$ symmetry. 
In our model, the fate of the accidental $Z_2$ symmetry is analogous to the fate of the accidental baryon number symmetry in the SM. The latter is broken by $D=6$ operators, leading to proton decay.  However, proton longevity  can be ensured 
by the details of the UV completion, giving either a long-lived or absolutely stable proton. Unsurprisingly, the situation is similar for our DM candidate. In this work we focus on the calculable renormalizable theory.

We  note that a number of earlier works have studied larger multiplets in
connection with neutrino
mass~\cite{Ahriche:2014oda,Babu:2009aq,Culjak:2015qja} (for related
phenomenology see Ref.~\cite{Hally:2012pu}). In particular,
Ref.~\cite{Culjak:2015qja} recently considered stable quintuplet
fermionic DM in a three-loop model of neutrino
mass.\footnote{Interestingly, the model of
Ref.~\cite{Culjak:2015qja} gives an accidental $Z_2$ symmetry after
imposing a separate $Z_2'$ symmetry.}


\section{Three-Loop Neutrino Mass and Lepton Flavor Violating Constraints\label{sec:nuetrino_mass7}}

The combination of the Yukawa Lagrangian and the terms
\begin{eqnarray}
V(H,S,\phi ) &\supset &\frac{\lambda _{\text{{\tiny S}}}}{4}(S^{-})^{2}\phi
_{abcdef}\phi _{a^{\prime }b^{\prime }c^{\prime }d^{\prime }e^{\prime
}f^{\prime }}\epsilon ^{aa^{\prime }}\epsilon ^{bb^{\prime }}\epsilon
^{cc^{\prime }}\epsilon ^{dd^{\prime }}\epsilon ^{ee^{\prime }}\epsilon
^{ff^{\prime }}+\mathrm{H.c.} \notag \\
&=&\frac{\lambda _{\text{{\tiny S}}}}{2}(S^{-})^{2}\{\phi ^{++++}\phi
^{--}-\phi ^{+++}\phi ^{-}+\phi ^{++}\phi ^{0}-\frac{1}{2}\phi ^{+}\phi
^{+}\}+\mathrm{H.c.}
\end{eqnarray}
in the scalar potential, are sufficient to explicitly break lepton number
symmetry. Consequently SM neutrinos are Majorana particles that acquire
radiative masses at the three-loop level, as shown in Figure~\ref%
{fig:7pletNuMass}. In the limit where the mass-splittings among
components of $\phi $ and $\mathcal{F}$ are neglected, the
calculation of the loop-diagram gives
\begin{equation}
(\mathcal{M}_{\nu })_{\alpha \beta }=\frac{7\lambda _{\text{{\tiny
S}}}}{(4\pi ^{2})^{3}}\frac{m_{\gamma }m_{\delta }}{M_{\phi
}}\,f_{\alpha \gamma }\,f_{\beta \delta }\,g_{\gamma i}^{\ast
}\,g_{\delta i}^{\ast }\times F\left( \frac{M_{i}^{2}}{M_{\phi
}^{2}},\frac{M_{\text{{\tiny S}}}^{2}}{M_{\phi
}^{2}}\right),\label{M_nu}
\end{equation}%
where the function $F$ encodes the loop integrals~\cite{JCAP-AN} and
$M_{\text{{\tiny S}}}$ is the charged-singlet mass.

\begin{figure}[t]
\begin{center}
\includegraphics[width = 0.5\textwidth]{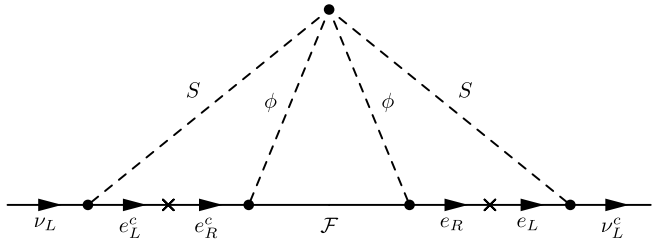}
\end{center}
\caption{Diagram for radiative neutrino mass, where $S$ and $\phi$
are new scalars and $\mathcal{F}$ is an exotic fermion. The lightest
component of $\mathcal{F}$ is a stable dark matter candidate.}
\label{fig:7pletNuMass}
\end{figure}

Neutrino masses calculated via Eq.~\eqref{M_nu} must satisfy the
data from neutrino oscillation experiments and reproduce the
following best-fit regions for the mixing angles and mass-squared
differences: $s_{12}^{2}=0.320_{-0.017}^{+0.016}$,
$s_{23}^{2}=0.43_{-0.03}^{+0.03}$,
$s_{13}^{2}=0.025_{-0.003}^{+0.003}$, $\Delta
m_{21}^{2}=7.62_{-0.19}^{+0.19}\times 10^{-5}\mathrm{eV}^{2}$, and
$|\Delta m_{13}^{2}|=2.55_{-0.09}^{+0.06}\times
10^{-3}\mathrm{eV}^{2}$~\cite{Tortola:2012te}. Matching to these
experimental values reveals the regions of parameter space where the
model gives viable neutrino masses.

The Yukawa couplings $g_{i\alpha}$ generate flavor changing
processes like $\mu\rightarrow e+\gamma$. Calculating the
corresponding diagrams in the limit where the mass-splittings are
neglected, and including the diagram containing the singlet $S$,
gives
\begin{eqnarray}
\mathcal{B}(\mu\rightarrow e\gamma) \simeq &\frac{\alpha v^4}{384
\pi} \times \left\{\frac{1764}{M_\phi^4}
\left|\sum_i g_{i\mu}g^*_{ie} F_2(M^2_{i}/M_{\phi}^2)\right|^2+\frac{%
|f_{\tau e}^*f_{\mu\tau}|^2}{M_\text{{\tiny S}}^4}\right\},
\label{eq:muEgamma_7}
\end{eqnarray}
where $F_2(x)=[1-6x+3x^2 +2x^3-6x^2\log x]/[6(x-1)^{4}]$ and $v$ is
the vacuum expectation value of $H$. The related expression for
$\mathcal{B}(\tau\rightarrow \mu+\gamma)$ is obtained by a simple
change of flavor labels in Eq.~\eqref{eq:muEgamma_7}. Replacing the
final-state electrons with muons in the diagram for $\mu\rightarrow
e+\gamma$ gives the one-loop contributions to the muon's anomalous
magnetic moment. In the limit where the radiative mass-splittings
are neglected these give
\begin{equation}
\delta a_{\mu }=-\frac{m_{\mu }^{2}}{16\pi ^{2}}\left\{ \sum_{i}\frac{%
7|g_{i\mu }|^{2}}{M_{\phi }^{2}}F_{2}(M_{i}^{2}/M_{\phi }^{2})+\sum_{\alpha
\neq \mu }\frac{|f_{\mu \alpha }|^{2}}{6M_{\text{{\tiny S}}}^{2}}\right\} .
\end{equation}%
The last term is due to the charged scalar $S$.

A further constraint of $(\mathcal{M}_\nu)_{ee}\lesssim0.35$~eV
follows from null-results in searches for neutrino-less double-beta
decay~\cite{Simkovic:2009pp}, though analysis shows that this
constraint is readily satisfied in the model. This constraint is
expected to improve after next generation
experiments, with an anticipated precision of $(\mathcal{M}%
_\nu)_{ee}\lesssim0.01$~eV~\cite{Avignone:2005cs}.

\section{Dark Matter\label{sec:dark_matter5}}

\subsection{Relic Density}

As mentioned already, the only viable DM candidate in the model is the lightest
neutral fermion $\mathcal{F}_{1}^{0}$. There are two classes of interactions
that can maintain thermal contact between the DM and the SM in the early
universe. Interactions mediated by the scalar $\phi $ have the cross section
\begin{equation}
\sigma (2\mathcal{F}^{0}\rightarrow \ell _{\beta }^{+}\ell _{\alpha }^{-})=%
\frac{|g_{1\beta }g_{1\alpha }^{\ast }|^{2}}{48\pi }\frac{M_{\text{{\tiny DM}%
}}^{2}(M_{\text{{\tiny DM}}}^{4}+M_{\phi }^{4})}{(M_{\text{{\tiny DM}}%
}^{2}+M_{\phi }^{2})^{4}}\times v_{r}\ \equiv \ \sigma _{0,0}^{\alpha \beta }
\label{eq:phi_neutral}
\end{equation}%
where $v_{r}$ is the DM relative velocity, in the centre-of-mass
frame. Note that there are no $s$-wave annihilations when
final-state lepton masses are neglected, as the DM is a Majorana
fermion. There are no coannihilations mediated by $\phi $, though
given the small radiative mass-splittings, one should include the
annihilations of singly-charged fermions:
\begin{equation}
\sigma (\mathcal{F}^{-}\mathcal{F}^{+}\rightarrow \ell _{\beta }^{+}\ell
_{\alpha }^{-})=\frac{|g_{1\beta }g_{1\alpha }^{\ast }|^{2}}{48\pi }\frac{M_{%
\text{{\tiny DM}}}^{2}(M_{\text{{\tiny DM}}}^{4}+M_{\phi }^{4})}{(M_{\text{%
{\tiny DM}}}^{2}+M_{\phi }^{2})^{4}}\times v_{r}\ \equiv \ \sigma
_{\pm }^{\alpha \beta }, \label{eq:phi_charged7}
\end{equation}%
and similarly for the higher-charged fermions
\begin{eqnarray}
\sigma (\mathcal{F}^{--}\mathcal{F}^{++}\rightarrow \ell _{\beta }^{+}\ell
_{\alpha }^{-}) &\equiv &\sigma _{\pm \pm }^{\alpha \beta }\ =\ \sigma _{\pm
}^{\alpha \beta }\,, \label{eq:phi_2charged7} \\
\sigma (\mathcal{F}^{---}\mathcal{F}^{+++}\rightarrow \ell _{\beta
}^{+}\ell _{\alpha }^{-}) &\equiv &\sigma _{\pm \pm \pm }^{\alpha
\beta }\ =\ \sigma _{\pm }^{\alpha \beta }. \label{eq:phi_3charged7}
\end{eqnarray}

There are also processes mediated by $SU(2)_{L}$ gauge bosons, which can be
calculated in the limit of an exact $SU(2)$ symmetry. The corresponding cross sections
 can be obtained with the results of Ref.~%
~\cite{Cirelli:2009uv}. Due to the small mass-splitting among the
components of $\mathcal{F}_{1}$, one should also include
coannihilation processes. Adding annihilation and coannihilation
channels together in the standard way gives~\cite{Griest:1990kh}
\begin{eqnarray}
&&\sigma _{eff}(2\mathcal{F}\rightarrow SM)\times v_{r} \notag \\
&=&\frac{1}{g_{eff}^{2}}\left[ \sigma _{W}\times
v_{r}+\sum_{\alpha,\beta }\left\{ g_{0}^{2}\,\sigma _{0,0}^{\alpha
\beta }+2g_{\pm }\,\sigma _{\pm }^{\alpha \beta }+2g_{\pm \pm
}\,\sigma _{\pm \pm }^{\alpha \beta }+2g_{\pm \pm \pm }\,\sigma
_{\pm \pm \pm }^{\alpha \beta }\right\} \times v_{r}\right],
\end{eqnarray}%
where the mass-splittings among fermion components are neglected and
the $SU(2)_L$ channels give
\begin{equation}
\sigma _{W}\equiv \frac{7\pi \alpha _{2}^{2}}{2M_{\text{{\tiny DM}}}^{2}v_{r}%
}\left\{ 1392+526v_{r}^{2}\right\} . \label{sigma_W}
\end{equation}%
In the above, $g_{eff}=g_{0}+2g_{\pm }+2g_{\pm \pm }+2g_{\pm \pm \pm }$,
with $g_{0}=g_{\pm }=g_{\pm \pm }=g_{\pm \pm \pm }=2$.

In principle one can calculate the mass range that gives a viable DM
relic density using the above expressions. However, the cross
section into gauge bosons
may be significantly enhanced by the non-perturbative Sommerfeld correction~%
~\cite{Hisano:2003ec, Hisano:2004ds, Hisano:2006nn}. One must solve
the Schr\"{o}dinger equation in terms of a non-relativistic bound
state of two DM particles in order to estimate the non-perturbative
Sommerfeld correction. The calculation is somewhat involved, though
the correction has been calculated for several $SU(2)_{L}$
multiplets in Ref.~\cite{Cirelli:2009uv} and the effect is found to
be important for larger multiplets. The enhancement of the cross
section influences the DM mass required to give the observed relic
density as the DM mass is the unique parameter that can control the
cross section when the annihilation cross section is dominated by
gauge interactions.\footnote{This is expected in the present model,
due to the relatively large value of $\sigma_{W}$ in
Eq.~\eqref{sigma_W}.} For example, the DM mass is shifted from $3.8$
\textrm{TeV} to $9.5$ \textrm{TeV} for a fermion quintuplet with
$Y=0$, from $5.0$ \textrm{TeV} to $9.4$ \textrm{TeV} for scalar
quintuplet with $Y=0$, and from $8.5~\mathrm{TeV}$ to
$25~\mathrm{TeV}$ for scalar septuplet with
$Y=0$~\cite{Farina:2013mla}. A similar enhancement is expected for
the fermion septuplet DM with $Y=0$ in our model, though a detailed
calculation is beyond the scope of this work. Guided by the results
listed in Ref.~\cite{Farina:2013mla} we expect the Sommerfeld
enhancement will increase the requisite DM mass by a factor of
approximately 3. As we shall see, this suggests the required DM mass
should be $\sim20-25$ \textrm{TeV} when the Sommerfeld effect is
taken into account.

The DM annihilation processes which induce monochromatic gamma-rays are also
enhanced by the Sommerfeld correction in the present universe.
This can be a significant signature of DM as an indirect detection signal.
Since the DM mass is predicted around $M_{\text{{\tiny DM}}}=$20$\sim$25 TeV
in our model, after including
Sommerfeld correction, monochromatic gamma-rays at
$E_\gamma=M_{\text{{\tiny DM}}}$ could be detected by future gamma-ray
experiments such as CTA~\cite{Bernlohr:2012we}.


\subsection{Direct Detection}

There is no tree-level coupling between DM and quarks. However, $W$
boson exchange gives three one-loop diagrams which can produce
signals at direct-detection experiments ~\cite{Ahriche:2014cda}.
There are both spin-dependent and spin-independent contributions to
the scattering, however, spin-dependent contributions are suppressed
by the heavy DM mass.
As we consider relatively heavy values of $M_{\text{{\tiny DM}}}>1$~\textrm{%
TeV}, the spin-dependent contributions can be neglected. Therefore
spin-independent scattering dominates and the cross section is determined by
SM interactions:
\begin{equation}
\sigma _{\mathrm{SI}}\,(\mathcal{F}^{0}N\rightarrow \mathcal{F}^{0}N)\simeq
\frac{36\pi \alpha _{2}^{4}M_{A}^{4}f^{2}}{M_{W}^{2}}\left[ \frac{1}{%
M_{h}^{2}}+\frac{1}{M_{W}^{2}}\right] ^{2}.
\end{equation}%
The DM scatters from a target nucleus $A$ of mass $M_{A}$, and the
standard parametrization for the nucleon is adopted:
\begin{equation}
\langle N|\sum_{q}m_{q}\,\bar{q}q\,|N\rangle \ =\ f\,m_{N}.
\end{equation}%
Here $m_{N}$ is the nucleon mass and $f=\sum_q f_q$ is
subject to the standard QCD uncertainties. For $f\approx0.3$, the cross section for the one-loop processes is $%
\sigma _{\mathrm{SI}}\simeq 4\times 10^{-44}\mathrm{cm}^{2}$, which
is just beyond the sensitivity of LUX~\cite{Akerib:2013tjd} for
heavy DM with $M_\dm\sim 25$ \textrm{TeV}. Note, however, that
recent lattice simulations suggest a somewhat lower value of strange
content $f_s\approx 0.043\pm0.011$~\cite{Junnarkar:2013ac}, which,
when combined with cancellations from two-loop diagrams, gives a
smaller cross section of $\sigma
_{\mathrm{SI}}\approx4\times10^{-46}\mathrm{cm}^{2}$~\cite{Hisano:2010ct}.\footnote{We
estimate the two-loop effect with a simple scaling of the results in
Ref.~\cite{Farina:2013mla}.} In either case, the result is beyond
the current sensitivity of LUX, though future discovery prospects
for the DM candidate can be considered promising.


\section{Numerical Results and Discussion\label{sec:results7}}

As already mentioned above, to determine the viable DM mass range
one should include the Sommerfeld enhancement. However, as a first
task we perform a numerical scan of the parameter space without the
Sommerfeld enhancement, determining the favored DM mass range. We
subsequently include a simple estimate of the effect.

\begin{figure}[t]
\begin{center}
\includegraphics[width=0.4\textwidth]{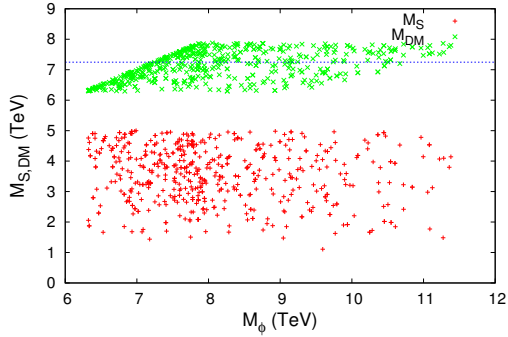}
\end{center}
\caption{The DM and charged scalar masses versus the scalar
septuplet mass for the case with no Sommerfeld enhancement. The blue
line at $M_{\dm}=7.2$ TeV gives the best-fit value for $\Omega
_{DM}h^{2}$ in the limit $g_{i\alpha }\rightarrow 0$.} \label{Omega}
\end{figure}

For the numerical scan we seek regions of parameter space that
satisfy the previously mentioned constraints, while simultaneously
giving neutrino masses and mixings in agreement with the
experimental values and a DM relic density within the range
$\Omega_{\text{{\tiny DM}}}h^{2}\sim 0.09-0.14$. We consider the
free parameter values
\begin{eqnarray}
\left\vert f_{\alpha \beta }\right\vert ^{2},\,\left\vert g_{i\alpha
}\right\vert ^{2} &\lesssim &9,\quad ~500~\mathrm{GeV}\leq M_{\text{{\tiny DM}}}\leq 10~\mathrm{TeV}, \notag \\
100~\mathrm{GeV} &\leq &M_{S}\leq 10~\mathrm{TeV},\quad
~M_{2,3},M_{\phi }\gtrsim M_{\text{{\tiny DM}}}. \label{Para}
\end{eqnarray}
The results for the values of $M_\dm$, $M_\s$ and $M_\phi$ are shown
in Figure \ref{Omega}. We find that viable neutrino masses can be
obtained for a large region of parameter space, though the DM mass
should be confined to the tidy range of 7.18-7.31~TeV for the relic
density to match the observed value. This region is somewhat tighter
than the corresponding region for the related models with
triplets~\cite{Ahriche:2014cda} and
quintuplets~\cite{Ahriche:2014oda}, due to the fact that the cross
sections for annihilations mediated by the couplings $g_{i\alpha}$,
namely Eqs.~\eqref{eq:phi_neutral}-\eqref{eq:phi_3charged7}, are
smaller compared to the contribution of $SU(2)_L$ gauge bosons
(\ref{sigma_W}). In the triplet and quintuplet cases
~\cite{Ahriche:2014cda, Ahriche:2014oda} the charged lepton
contribution is non-negligible, allowing a greater spread for the DM
mass interval.

The Sommerfeld enhancement is expected to increase the required DM
mass by a factor of roughly 3. Therefore, in order to approximately
take this effect into account, we redo the numerical scan with the
DM mass in the relic density replaced by $M_{\text{{\tiny DM}}}/3$,
searching for parameter space that gives viable
 neutrino masses and mixings and is consistent with low-energy constraints. This approach only provides a rough approximation for the value of the DM mass but, importantly, it allows us to discover if the requisite heavier values of $M_\dm$ and $M_\phi$ are compatible with the low-energy data. Note that, because the relic density calculation has a reduced sensitivity to the couplings $g_{i\alpha}$ (as DM annihilations in the early universe are dominated by $SU(2)_L$ annihilations), the key question is whether there is viable parameter space that achieves neutrino mass and satisfies the constraints, given the heaviness of the DM. Our approach allows us to answer this question and a small shift in $M_\dm$ should not significantly affect the conclusion.

\begin{figure}[t]
\begin{center}
\includegraphics[width=5.5cm,height=5cm]{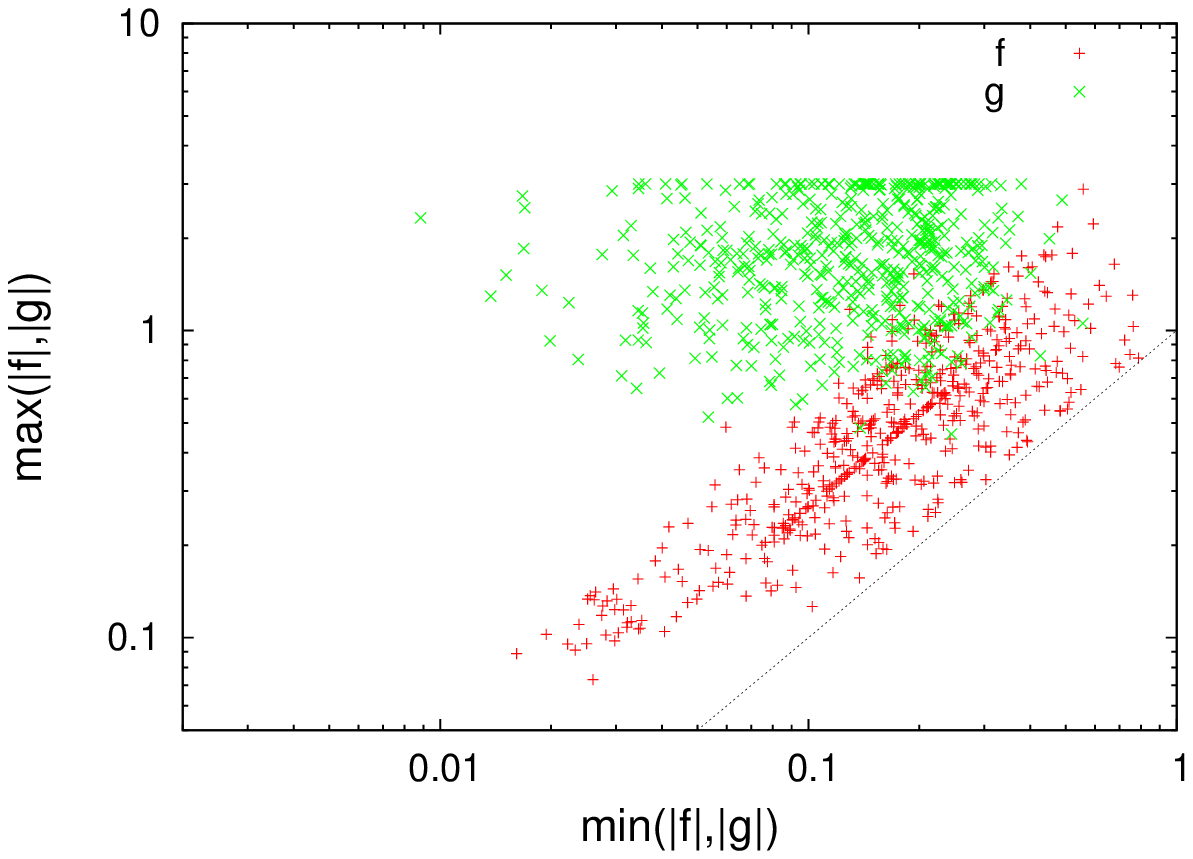}~
\includegraphics[width=5.5cm,height=5cm]{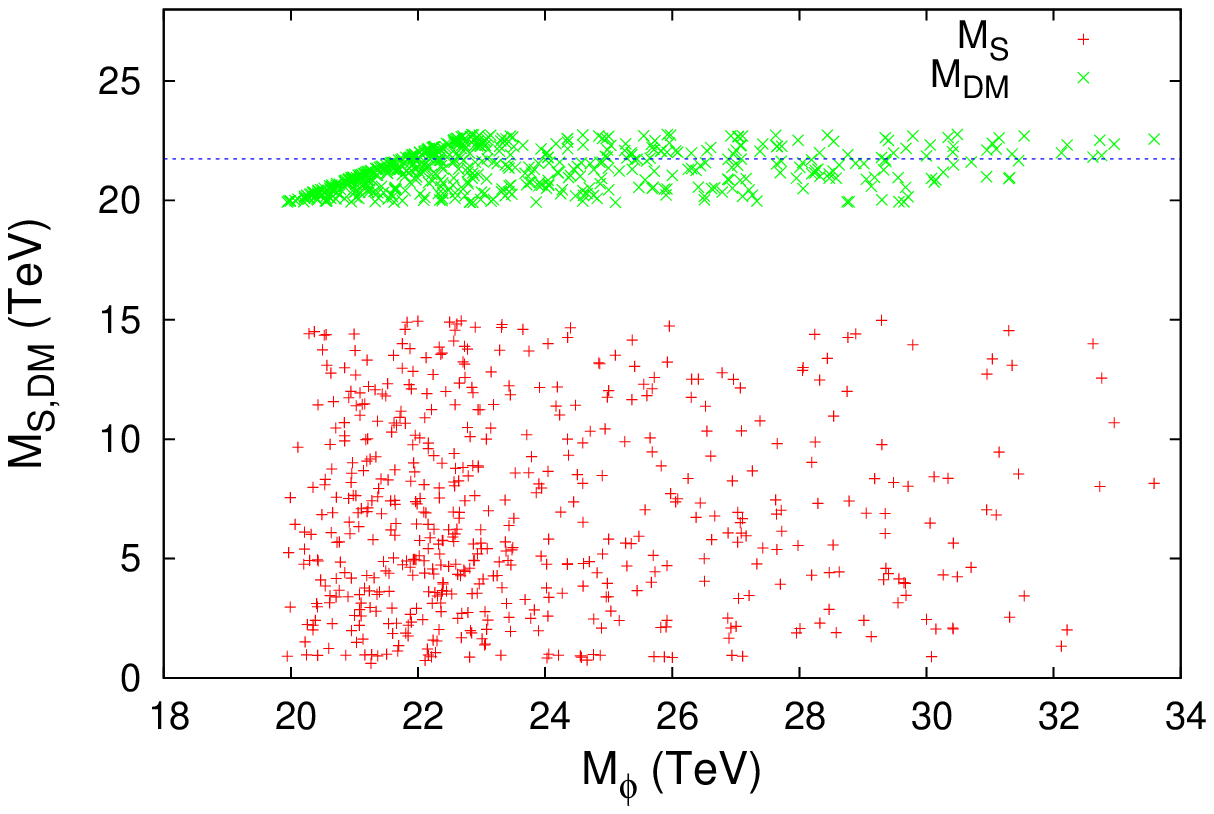}~
\includegraphics[width=5.5cm,height=5cm]{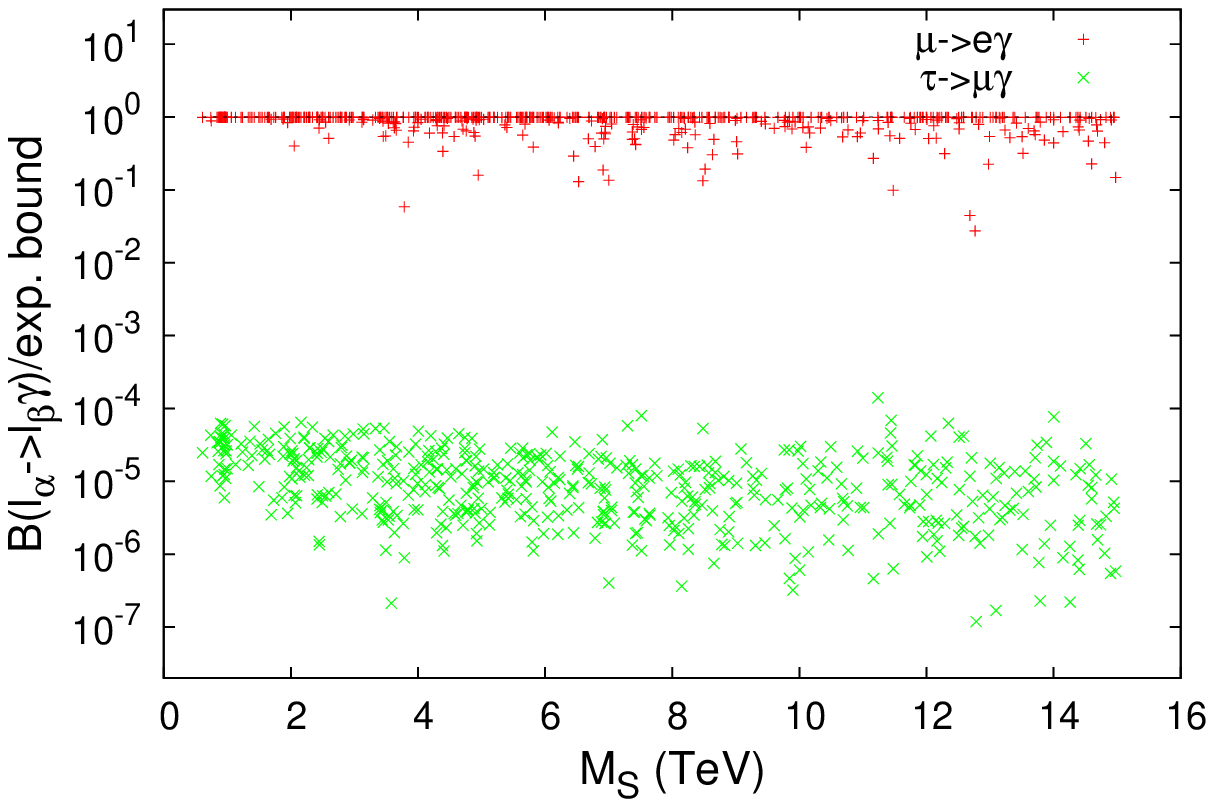}
\end{center}
\caption{Numerical results for the case where the Sommerfeld
enhancement is considered. Left: the Yukawa coupling range that
satisfies all previous constraints. The dashed line represents the
case where all couplings are close in absolute value, i.e.,
$min(|f|)=max(|f|)$. Center: the masses $M_\dm$ and $M_\s$ versus
the septuplet scalar mass $M_\phi$. The blue line gives the $M_\dm$
best-fit value for $\Omega _{DM}h^{2}$ when $g_{i\alpha }\rightarrow
0$. Right: The constraints from lepton flavor violating decays
scaled by the experimental bounds versus the charged scalar mass.
Here the muon anomalous magnetic moment is smaller than the
experimental bound by more than one order of magnitude.} \label{Oms}
\end{figure}

Performing the modified numerical scan produces the new results
shown in Figure \ref{Oms}. There is considerable parameter space
that satisfies the constraints with the DM mass in the range
19.7-23.1~\textrm{TeV}, centered around the value of
$M_{\text{{\tiny DM}}}=21.7$~\textrm{TeV}, which is preferred in the
limit $g_{i\alpha}=0$. The scalar $\phi$ must now be heavier than
19.9 \textrm{TeV}, while the charged scalar singlet $S$ can remain
as light $\sim500$~GeV, similar to the case without Sommerfeld
enhancement effect. One observes that the branching ratio $B(\tau
\rightarrow \mu +\gamma )$ is smaller than the experimental bound by
4-6 orders of magnitude while the constraint of $B(\mu \rightarrow
e+\gamma )<5.7\times 10^{-13}$ is more severe. In particular, it is
evident that improved measurements of $B(\mu \rightarrow e+\gamma )$
are capable of excluding the model. Though not shown in the figure,
the preferred regions of parameter space are not ruled out by the
data on the anomalous magnetic moment of the muon; the extra
contribution from the exotics can contribute to the observed
discrepancy, though it cannot explain it
entirely~\cite{Bennett:2006fi, Jegerlehner:2009ry}.

We note that with only two generations of fermions $\mathcal{F}_{i}$
($g_{3\alpha }=0$), the bound on $B(\mu \rightarrow e+\gamma )$ is
violated. Therefore three generations of $\mathcal{F}_{i}$ are
required to remain consistent with constraints from lepton flavor
violating processes. Also, the neutrino data prefers that one does
not introduce large hierarchies between $M_\dm$ and the other exotic
masses, $M_{2,3}$ and $M_\phi$, with
$M_{\phi,2,3}\sim\mathcal{O}(1-10)\times M_\dm$ preferred. The
exotics are therefore clustered near $M_\dm$. Finally, we emphasize
that the preferred values of $M_\dm$ should only be taken as a
guide, though our analysis clearly shows that one can satisfy the
low-energy constraints with the required heavier values of $M_\dm$
and $M_\phi$.


\section{Conclusion\label{sec:conc7}}

We presented an original model of radiative neutrino mass that \emph{%
automatically} contains an accidental $Z_{2}$ symmetry and thus
provides a stable DM candidate. This gives a common description for
neutrino mass and DM without invoking any symmetries beyond those
present in the SM. The DM is the neutral component of a septuplet
fermion $\mathcal{F}\sim (1,7,0)$, and should have mass
$M_{\text{{\tiny DM}}}\approx 20-25$~\textrm{TeV}. The model can
give observable signals via flavor-changing leptonic decays and DM
direct-detection experiments. It also predicts a charged scalar $S$
that can be at the \textrm{TeV} scale and within reach of future
colliders.

\section*{Acknowledgments\label{sec:ackn}}

AA is supported by the Algerian Ministry of Higher Education and
Scientific Research under the CNEPRU Project No. D01720130042. KM is
supported by the Australian Research Council. TT acknowledges
support from P2IO Excellence Laboratory.

\end{document}